\documentclass[preprint,review,12pt]{elsarticle}
\usepackage{graphics, amsmath, amsthm, amssymb}
\usepackage{epsfig,multirow,multicol}
\usepackage[colorlinks]{hyperref}
\newcommand{\be}{\begin{equation}}
\newcommand{\ee}{\end{equation}}
\newcommand{\bea}{\begin{eqnarray}}
\newcommand{\eea}{\end{eqnarray}}
\newcommand{\bwt}{\begin{widetext}}
\newcommand{\ewt}{\end{widetext}}

\newcommand{\bff}{\bf}
\journal{Journal of Magnetism and Magnetic Materials}

\begin{document}

\begin{frontmatter}



\title{Thermodynamic properties of ferroics described by the transverse Ising model and their applications for CoNb$_2$O$_6$}
\author{Cong Thanh Bach, Niem Tu Nguyen and Giang Huong Bach}
\address{Faculty of Physics, VNU University of Science, 334 Nguyen Trai, Hanoi, Vietnam}


\begin{abstract}
	The temperature and transverse field dependence of the entropy and the specific heat of ferroics in the quantum paramagnetic (QPa) state is investigated using the transverse Ising model (TIM) with different spin values within mean field and Gaussian spin fluctuation approximations.
  A maximum peak of the temperature dependence of the specific heat curves is enhanced in the QPa state due to spin fluctuations. This peak shifts to higher temperature region and its magnitude reduces with increasing transverse field. In addition, the temperature corresponding to this maximum depends linearly on the deviation of the transverse field from its critical value. The obtained specific heat qualitatively agrees with the experimental observation for the quasi-one-dimensional (1D) Ising ferromagnet CoNb$_2$O$_6$ in the QPa phase. It is also shown that the  spin-1/2 three dimensional (3D) TIM clearly describes the specific heat of CoNb$_2$O$_6$ in the QPa states near the critical temperature. However, the spin-3/2 3D-TIM is more adequate than the spin-1/2 3D-TIM for describing the thermal behavior of CoNb$_2$O$_6$ in the QPa states at high fields and at elevated temperatures.

\end{abstract}


\end{frontmatter}


\section{Introduction}

Thermodynamic characteristics such as entropy, heat capacity or susceptibility of the ferroic materials (ferromagnets or ferroelectrics) having quantum phase transitions (QPT) are studied intensively to comprehend their specific nature. Quantum criticality near the quantum critical point (QCP) observed firstly in ferroelectrics where the ferroelectric transition temperature is suppressed to zero by tuning parameters \cite{Scott2015}. The tuning parameters, which control the systems from ordered phase to quantum para-magnetic (QPa) or quantum para-electric states in different cases, can be hydro-static pressure, atomic substitution, a transverse magnetic field in ferromagnets or an electric field in ferroelectrics. 

The non-classical behavior of the inverse susceptibility $\chi^{-1}$ (equivalent to the inverse dielectric function $\epsilon^{-1}$ ) of SrTiO$_3$ and related compounds is proportional to $T^{2}$ close to the QCP \cite{Rowley2014}. In order to convince of the weight of the finite temperature quantum criticality \cite{Sachdev1999}, Kinross et al. have specified the quantum critical properties of a quasi-one-dimensional Ising ferromagnet CoNb$_2$O$_6$ sustaining up to high temperature. The temperature is $\rm T\approx 4J$ with J the exchange interaction between nearest neighbor spins \cite{Kinross2014}. Besides, Tian Lang et al. \cite{Tian2015} have observed a prominent peak in the heat capacity curve of the CoNb$_2$O$_6$ at the QCP. They have applied the exact solution \cite{Lieb1961,Pfeuty1970} for the 1D transverse Ising model with spin $s=1/2$ to explain the existence of the prominent peak and to provide the evidence for the gap-less fermion-like excitation in a narrow interval of the transverse magnetic field below the QCP. 

Even though those results are interesting, their explanation is not unique and needs more discussions. The anisotropy of the heat capacity and of the susceptibility of CoNb$_2$O$_6$ has also been investigated experimentally since 1994 by Hanawa et al. \cite{Hanawa1994}. They confirm that the magnetic moment of Co in this compound is 5.05 $\mu_B$ and Co$^{+2}$ (3d$^7$) ion prefers a high spin (HS) state with s=3/2  rather than a low spin (LS) state with s=1/2 which is normally used in the exact fermion solution for the TIM. Recent first-principle calculations of Molla and Rahaman \cite{Molla2018} have indicated that the magnetic moment at cobalt site is 2.89 $\mu_B$, thus cobalt ion favors a HS state. What is the relevance of the HS model for the description of the thermodynamics of the spin system in the varying transverse field (TrF)? The specific heat of CoNb$_2$O$_6$ in the QPa states (Fig.~5 of Ref.~[5]) clearly exhibits the maximum peak which gradually reduces and moves to higher temperatures while increasing field. A big discrepancy between the experimental data and the theoretical QPa specific heat curves given by the exact fermion solution for CoNb$_2$O$_6$ requires additional investigations.

In a previous work \cite{Cong2018}, we have shown the existence of the gap-less long-wavelength spin excitations at the QCP of the mono-spin layer using the XZ quantum Heisenberg model with an arbitrary spin value under the influence of the TrF. It implies that there is another way to interpret the finite temperature experimental results using the TIM with different spin values beside the famous spin-$1/2$ TIM given in the literature (see, for example, Ref.~\cite{Suzuki2013}). A coupling between spin chains, which plays an essential role in the formation of isosceles triangular lattice planes of CoNb$_2$O$_6$, is essentially taken into account in the 3D spin model \cite{Cabrera2014}. The specific heat in the transverse Ising thin films has been studied in Ref.~\cite{Kane2004} using the mean field (MF) and the effective mean field theory but the TrF dependence of the specific heat has not been considered yet.

In this paper, we use the TIM with different spin values to describe the temperature and the transverse field dependence of the entropy and of the specific heat of ferroics. Our calculations are performed within MF and Gaussian spin fluctuation approximations beyond the critical region. We also focus on the thermodynamic properties in the QPa states of ferromagnets and compare our results with the experimental specific heat of CoNb$_2$O$_6$. 

Our paper includes four sections. In section 2, the expressions of thermodynamic quantities within the MF and Gaussian approximations in the QPa states are given explicitly using the TIM model with arbitrary spin values.  Section 3 presents a comparison with the specific heat experiment for CoNb$_2$O$_6$ in the QPa regime and gives detail discussions. A conclusion is provided in the last section. Throughout section 2, we use a natural unit system with $\hbar=1$ and $ \rm k_B =1$. 

\section{The transverse Ising model and thermodynamic quantities}

\subsection{Model and free energy calculation in the Gaussian approximation}

The crystal structure of CoNb$_2$O$_6$ belongs to the space group $Pbcn$. The lattice parameters of an  orthorhombic unit cell of CoNb$_2$O$_6$ are $\rm a=14.1337$ $\AA$, $\rm b=5.7019$ $\AA$ and $\rm c=5.0382$ $\AA$ \cite{Heid1995}. To describe the magnetic behavior of the quasi-1D magnet CoNb$_2$O$_6$, we use a three-dimensional (3D) TIM instead of the quasi-2D model of Ref.~\cite{Cong2016}. The three basis vectors of the unit cell of the 3D spin lattice are chosen similarly to Ref.~\cite{Cabrera2014} where $\bf{a}_1= \bf{b}$, $\bf{a}_2=(\bf{a-b})/2$, $ \bf{a}_3= \bf{c}/2$. A spin position is defined by a three-component spin lattice vector $\bf R_j$. The z-axis of the crystallographic coordinate system Oxyz is parallel to the $\bf c$ vector and the external TrF directs along the x-axis which is parallel to the $\bf b$ vector. Denoting $\rm s^z_j$, $\rm s^x_j$ as spin operator components on the Oz, Ox coordinate axes, the Hamiltonian of the TIM is written by 
\be
\rm H = -h_0\sum_j s^{z}_{j} - \Omega_0 \sum_j s^{x}_{j} - {1 \over 2} \sum_{jj'} J_{jj'} s^{z}_{j} s^{z}_{j'}.
\label{Eq1a}
\ee
Here the external longitudinal h$_0$, the transverse field $\Omega_0$ and the exchange interaction $\rm J_{jj'}=\rm J(\bf |{R_j-R_{j'}}|)$ have the energy dimension. Separating H into mean field H$_0$ and spin fluctuation H$\rm _{int}$ parts and using a unitary rotation to transform spin operators $\rm s^z_j$, $\rm s^x_j$ to $\rm S^z_j$, $\rm S^x_j$ in the new  coordinate system OXYZ \cite{Cong2018}, we obtain a transformed Hamiltonian of the spin system, $\rm H= \rm H_0 +H_{int}$, where
\bea
\rm H_0&=&\rm{N\over 2} J(0) m^2_z -\gamma \sum_{j} S^{Z}_j, \\
\rm H_{int} &=& -{1\over 2} \sum_{{\bf k},\alpha \alpha'} \rm I^{\alpha \alpha'}({\bf k}) \delta S^{\alpha}({\bf k})\delta S^{\alpha'}(-{\bf k}),
\label{Eq3}
\eea  
with $\alpha= \rm X,Z$. $\rm J(0)$ is a Fourier component of exchange interaction at $\rm \bf k=0$. $\rm m_z=\rm \langle s^z \rangle$ and $\rm m_x=\langle s^x \rangle$ are the thermodynamic average of the magnetic moments per site. Since the OZ axis of the rotated coordinate system OXYZ is chosen parallel to the direction of the total field $\gamma$, only a statistical average value of the longitudinal spin component differs from zero ($\rm \langle S^z_j \rangle \neq 0$ and $\rm \langle S^x_j \rangle=0$). The total field $\gamma=\sqrt{\rm h^2+\Omega^2_0}$ contains the longitudinal $\rm h=h_0+J(0)m_z$ and the transverse  $\Omega_0$ components. $\delta \rm S^{\alpha}({\bf k})$ is the Fourier image of the spin fluctuation operator. The symmetric 2x2 matrix of exchange interaction $\rm \hat{I} ({\bf k})$ with matrix elements appeared in Eq.~\ref{Eq3} is defined by
\bea
\rm \hat{I}({\bf k}) =
\begin{bmatrix}
	\rm I^{XX}({\bf k}) & \rm I^{XZ}({\bf k}) \\
	\rm I^{ZX}({\bf k}) & \rm I^{ZZ}({\bf k}) 
\end{bmatrix} 
= {J({\bf k}) \over {\gamma^2}}
\begin{bmatrix}
	\Omega^2_0 & -\Omega_0 \rm {h} \\
	-\Omega_0 \rm{h} & \rm h^2 
\end{bmatrix},
\label{Eq4}
\eea
with the Fourier image of the exchange interaction $\rm{J}(\bff{ k})=\sum_{R_j} J(R_j) e^{i\bff{kR_j}} $. The lattice vector $\bff{R_j}$ presented by three basis vectors $\bf a_1$, $\bf a_2$, $\bf a_3$ points out the position of the $\rm j^{th}$ spin. The wave vector $\bf k$ is given in the reciprocal lattice of the orthorhombic crystal lattice. Using the intra-chain and the inter-chain nearest (NN), next-nearest neighbor (NNN) exchange couplings for CoNb$_2$O$_6$ similarly to those in Ref.~\cite{Cabrera2014}, the Fourier image $\rm J(\bf k)$ is written in the following form, 
\bea
\rm J({\bf k})&=&\sum_{\bf \Delta} \rm{J'_z} e^{i\bf k \Delta} + \sum_{\bf \Delta_1} J_ze^{i \bf k \Delta_1} +  \sum_{\bf \Delta_2} J_1e^{i\bf k\Delta_2} +  \sum_{\bf \Delta_3} J_2e^{i \bf k\Delta_3} , \,\,\,\,\,\,\,\, \\
\bf \Delta&=&\pm \bf{c}; \Delta_1=\pm \bf{c}/2; \Delta_2=\pm \bf{b}; \Delta_3=\pm (\bf{a\pm b})/2. \,\,\,\,\,\,\,\,\,
\eea
Here J$_z>0$ (J'$_z<0$) is a ferromagnetic-FM (anti-ferromagnetic-AF) intra-chain exchange coupling between NN (NNN) spins along the {\bf z} direction of the crystallographic coordinate system. J$_1$ and J$_2$ are the inter-chain anti-ferromagnetic NN and NNN exchange coupling between spins in the xy-plane, respectively. In the next part of section 2, the Fourier image of the exchange coupling J$(\rm{\bf k})$, field strengths $\gamma$, h, $\Omega_0$, temperature $\tau$, free energy and spin wave frequency are given in terms of the exchange coupling J$_z$. For example, $\tau=\rm{T/J_z}$ and
\bea
\rm J({\bf k})&=&2[\rm{cos(k_zc/2)+J'_zcos(k_zc)+J_1cos(k_yb)} \nonumber\\ 
&+&\rm{2J_2cos(k_xa/2)cos(k_yb/2)]},\\
\rm{J(0)}&=&\rm{2(1+J'_z+J_1+2J_2)}.
\label{Eq8}
\eea 

Since we are interested in the role of the TrF, the external longitudinal field is turned off, h$_0=0$ and we have only the intrinsic longitudinal field, $\rm h=\rm J(0)m_z$. Using the functional integral method, the MF and Gaussian approximations similarly to Ref.~\cite{Cong2018}, we obtain the free energy per spin, $f = f_0+f_1$, including the mean field f$_0$ and the fluctuation f$_1$ parts, which are
\bea
\rm {f}_0&=& {1 \over 2}\rm{J(0)m}^2_z - {1\over \beta} \rm{ln}{ {\rm{sh[(s+1/2)y]} \over {sh(y/2)}} },  \label{Eq9} \\
\rm{f}_1&=& {1 \over {2\rm{N}\beta}} \sum_{\bf k}\rm{ln}\Big \{{1-{\beta I^{ZZ}({\bf k})\gamma b'_s(y) \over {\omega(\bf k)}}} \Big \} \nonumber \\
&+& {1 \over {2\beta \rm N}} \sum_{\bf k} \rm{ln}\Big  \{ \rm{sh}[{{\beta \omega(\bf k)} \over 2}]/\rm{sh}({y \over 2})\Big \},\\
\rm{y} &=& \beta \gamma ,
\label{Eq10}
\eea  
with $\beta=\tau^{-1}$. The temperature-dependent energy of the elementary excitation obtained in Ref.~\cite{Cong2018} is $\omega({\bf k})=\gamma-\rm{I}^{XX}({\bf k})b_s(\rm{y})$. $\rm b_s(y)$ and $\rm b'_s(y)$ are the Brillouin function and its derivative respectively,
	\bea
	{\rm b_s(y)} &=& \rm {(s+{1\over 2})cth \Big[ \Big( s+{1\over 2}\Big) y \Big]}- { 1 \over 2} cth {y\over2},\\
	\rm b'_s(y) &=& \rm {1 \over 4sh^2 \Big( {y\over2} \Big)} -{(s+{1 \over 2})^2 \over sh^2 [ (s+{1\over2})y ]}.
	\eea 

\subsection{Thermodynamic quantities of the TIM in the mean field approximation}

\begin{figure*}[tp]
	\centering
	\includegraphics*[height=2.5in,width=2.4in]{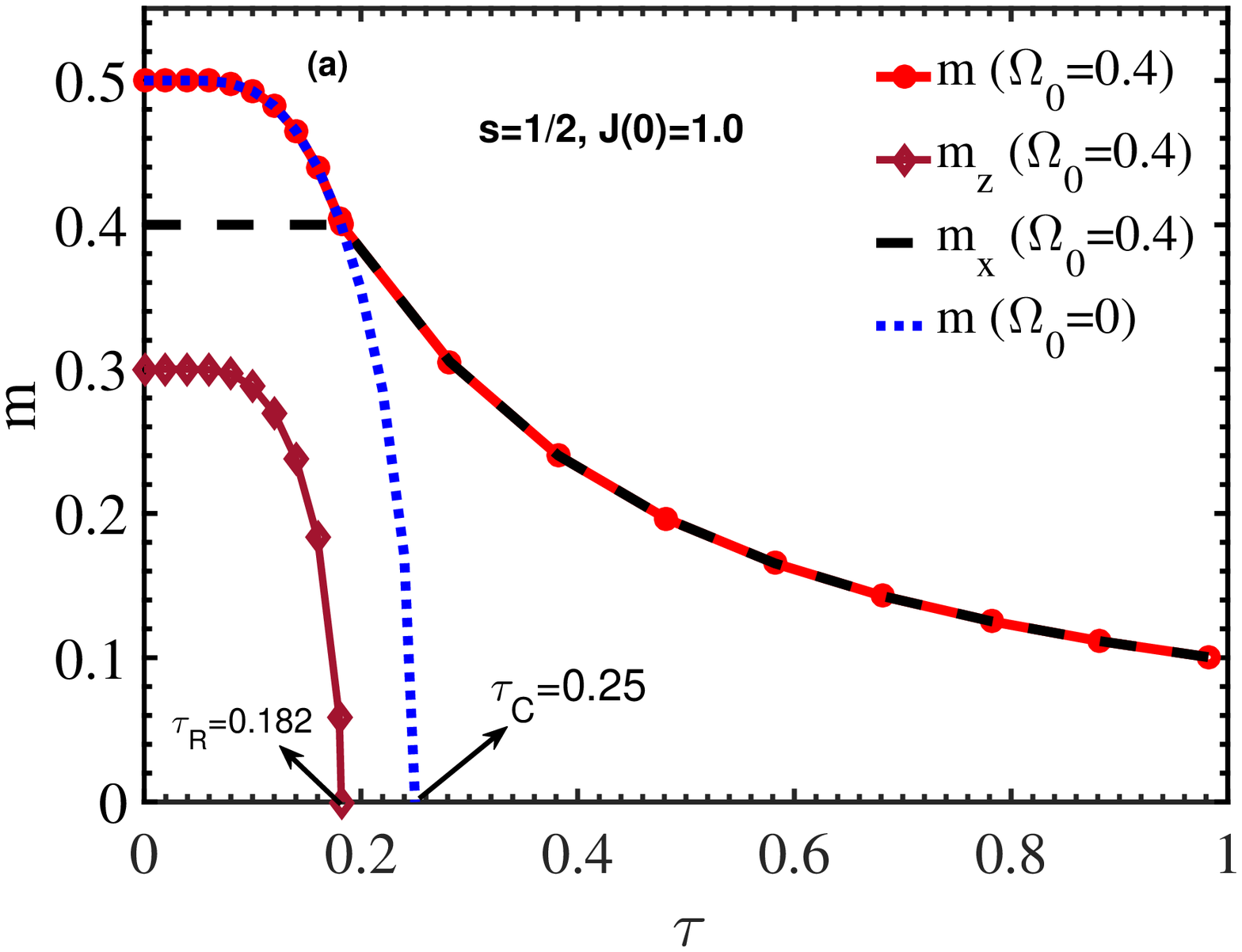} 
	\includegraphics*[height=2.6in,width=2.5in]{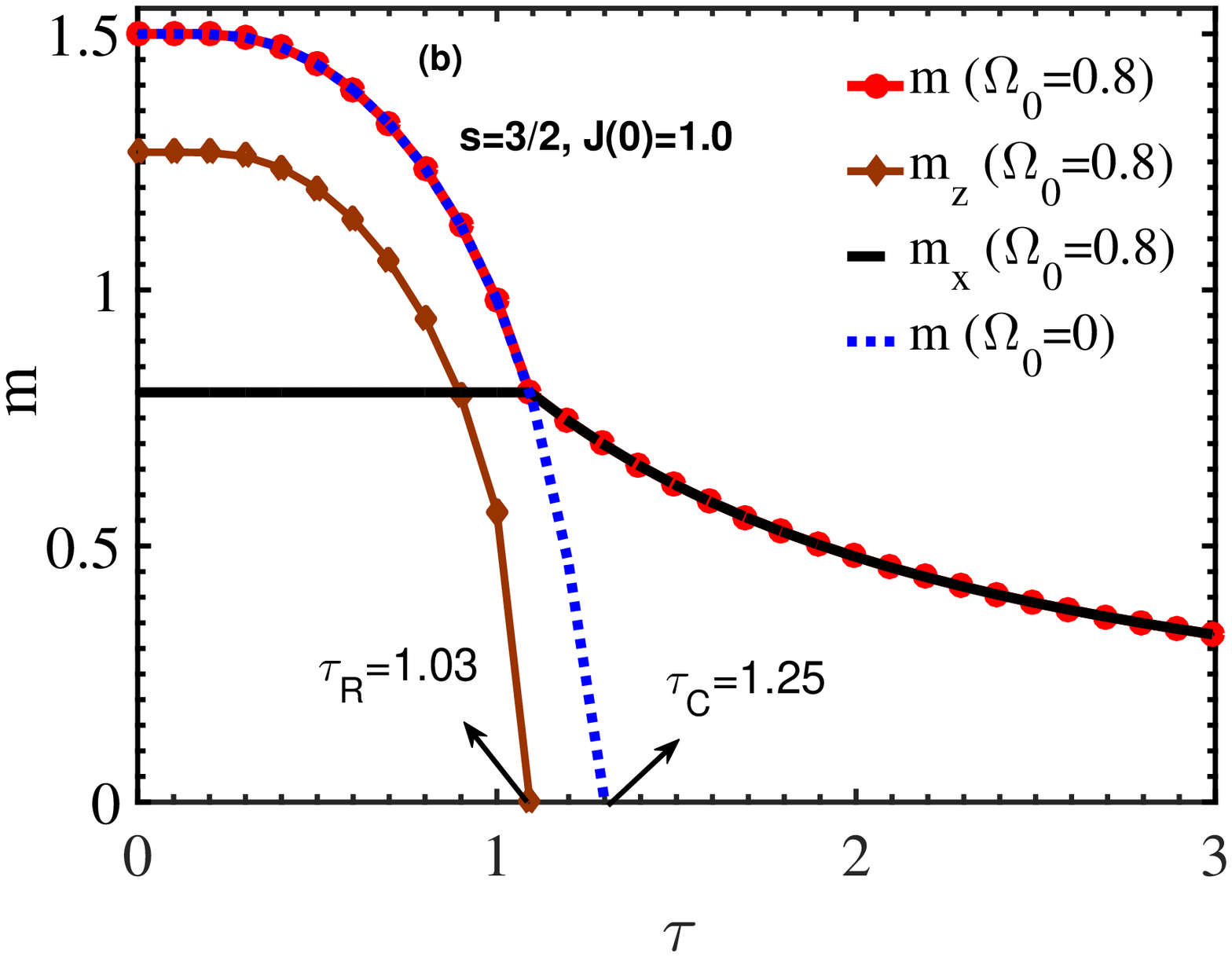} 
	\caption{The temperature dependence of the components m$\rm_x$, m$\rm_z$ and the total  magnetic moment per site m for spin (a) $\rm s=1/2$ and (b) $\rm s=3/2$ cases with $\rm J(0)=1.0$ at different transverse fields. The arrows indicate the spin reorientation temperature $\rm \tau_R$ and the Curie temperature $\rm \tau_C$. The parameters J(0), $\Omega_0$, $\rm \tau_R$, $\rm \tau_C$ are given in terms of NN exchange integral J$\rm _z$.}
	\label{Fig1}	
\end{figure*}
\begin{figure}[tp]
	\centering
	\includegraphics[height=2.4in,width=2.6in]{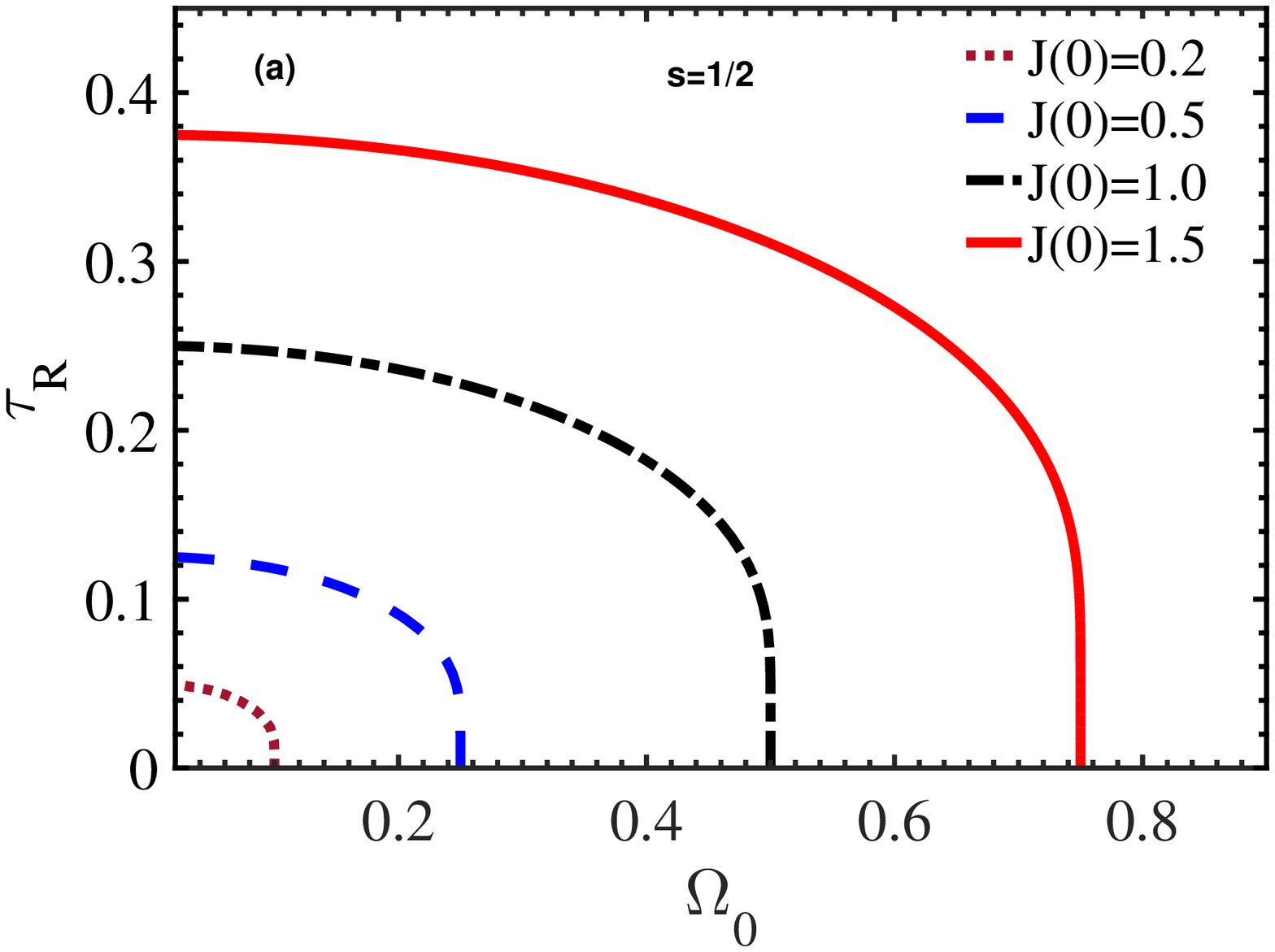} 
	\includegraphics[height=2.4in,width=2.6in]{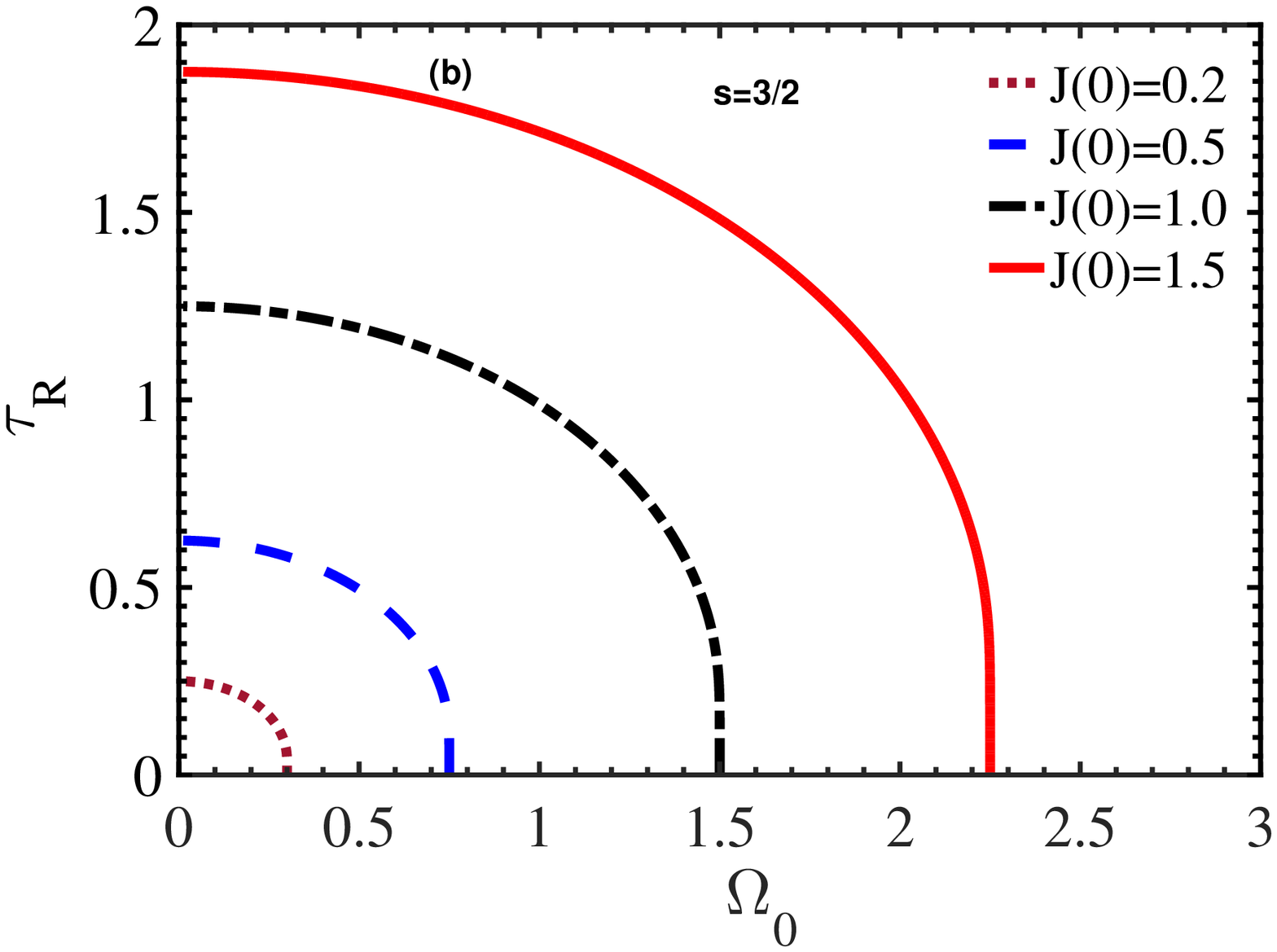} 
	\caption{The dependence of the spin reorientation temperature $\rm \tau_R$ on the transverse field $\Omega_0$ for spin (a) $\rm s=1/2$ and (b) $\rm s=3/2$ cases.}	
\end{figure}

Finite temperature behaviors of the spin system are described using the mean field approximation (MFA), where the spin fluctuation term $\rm{H_{int}}$ in Eq.~(\ref{Eq3}) is omitted.
The entropy S$_0$, the internal energy E$_0$ and the specific heat C$_0$ per spin derived by taking derivatives of Eq.~(\ref{Eq9}) with respect to temperature respectively are 
\bea
\rm{S_0}&=&-{\rm {\partial f_0} \over {\partial \tau} }=\rm -yb_s(y) + ln {{sh[(s+1/2)y]} \over {sh(y/2)}},\\
\rm{E_0}&=& \rm {f_0} + \tau \rm{S_0} = {1 \over 2} \rm {J(0) m^2_z} - \tau \rm{y b_s(y)},\\
\rm{C_0}&=& -\tau {\partial \rm S_0 \over \partial \tau} = \rm {y^2b_s'(y)} \Big\{ 1- {\rm{J(0)h^2 y b_s'(y)} \over \gamma^3 -  \rm{J(0)} \Omega^2_0 \rm{b_s(y)}} \Big\}^{-1}.\,\,\,\,\,\,\,\,
\eea 
The critical temperature $\rm \tau_R$ or the spin reorientation temperature \cite{Cong2018} is found by solving the following equation,
\be
\rm b_s \Big( {\Omega_0 \over \tau_R}\Big)= {\Omega_0 \over J(0)}.
\label{Eq20}
\ee
If $\tau_R(\Omega_{0c})=0$, the critical field deduced from Eq.~(\ref{Eq20}) is 
\be
\rm \Omega_{0c}=sJ(0).
\label{Eq21}
\ee
The magnetization components satisfy the following equations at different temperature regions.

i/ $\tau <\rm \tau_R$,
\be
\rm m_x = {\Omega_0 \over J(0)}; \,\,\,\, 
m_z= \sqrt{b^2_s(\gamma/\tau) - m^2_x}\,,
\ee
where $\gamma$ is the solution of the equation $\gamma=\rm J(0)b_s(\gamma/\tau)$ at a given temperature $\tau$.

ii/ $\rm \tau \ge \rm \tau_R$, 
\be
\rm m_z=0; \,\,\,\,
m_x = b_s(\gamma/\tau)\,,
\label{Eq23}
\ee
where $\gamma =\Omega_0$.

\begin{figure}[tp]
	\centering
	\includegraphics[height=2.4in,width=2.6in]{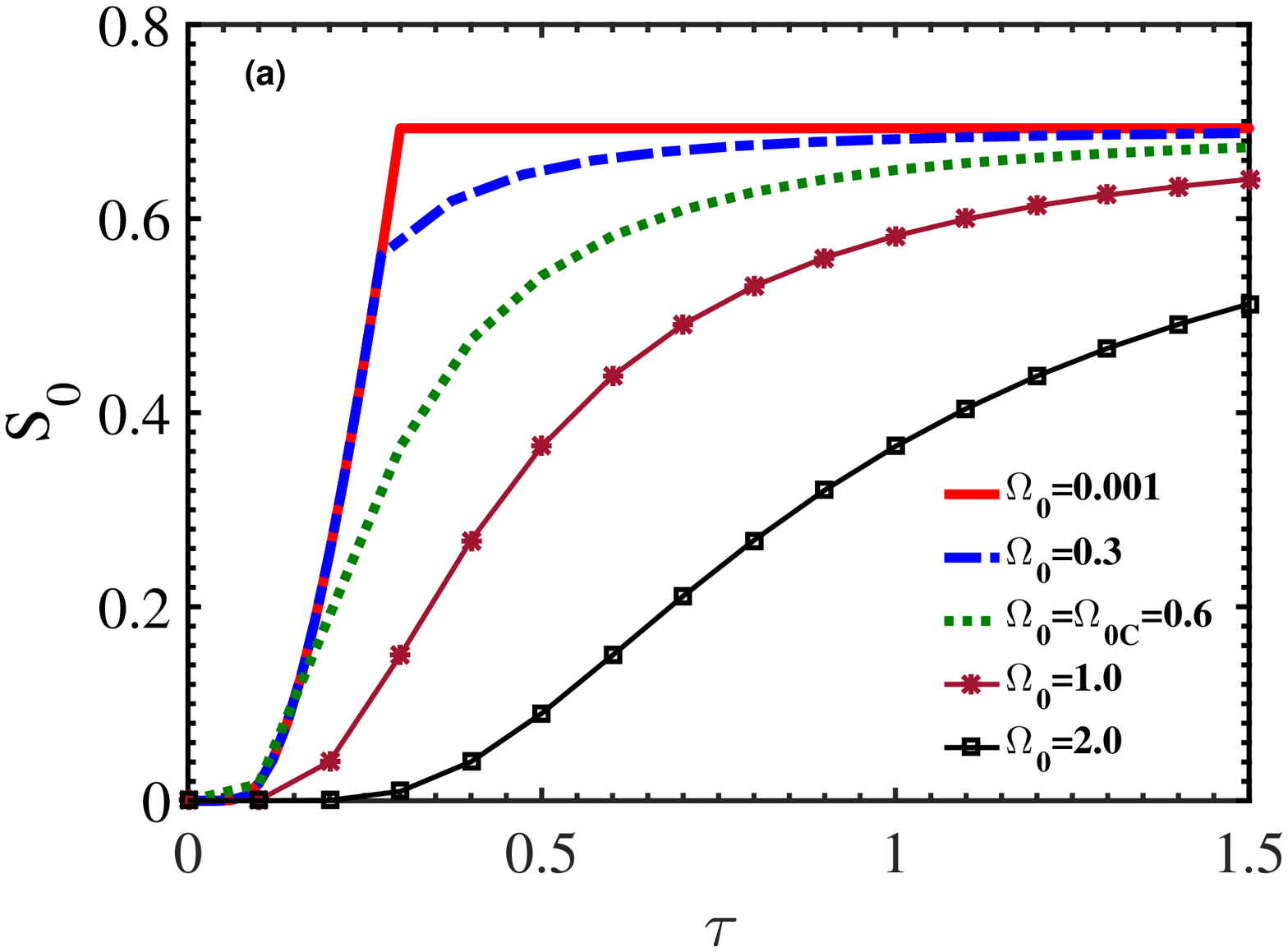} 
	\includegraphics[height=2.4in,width=2.6in]{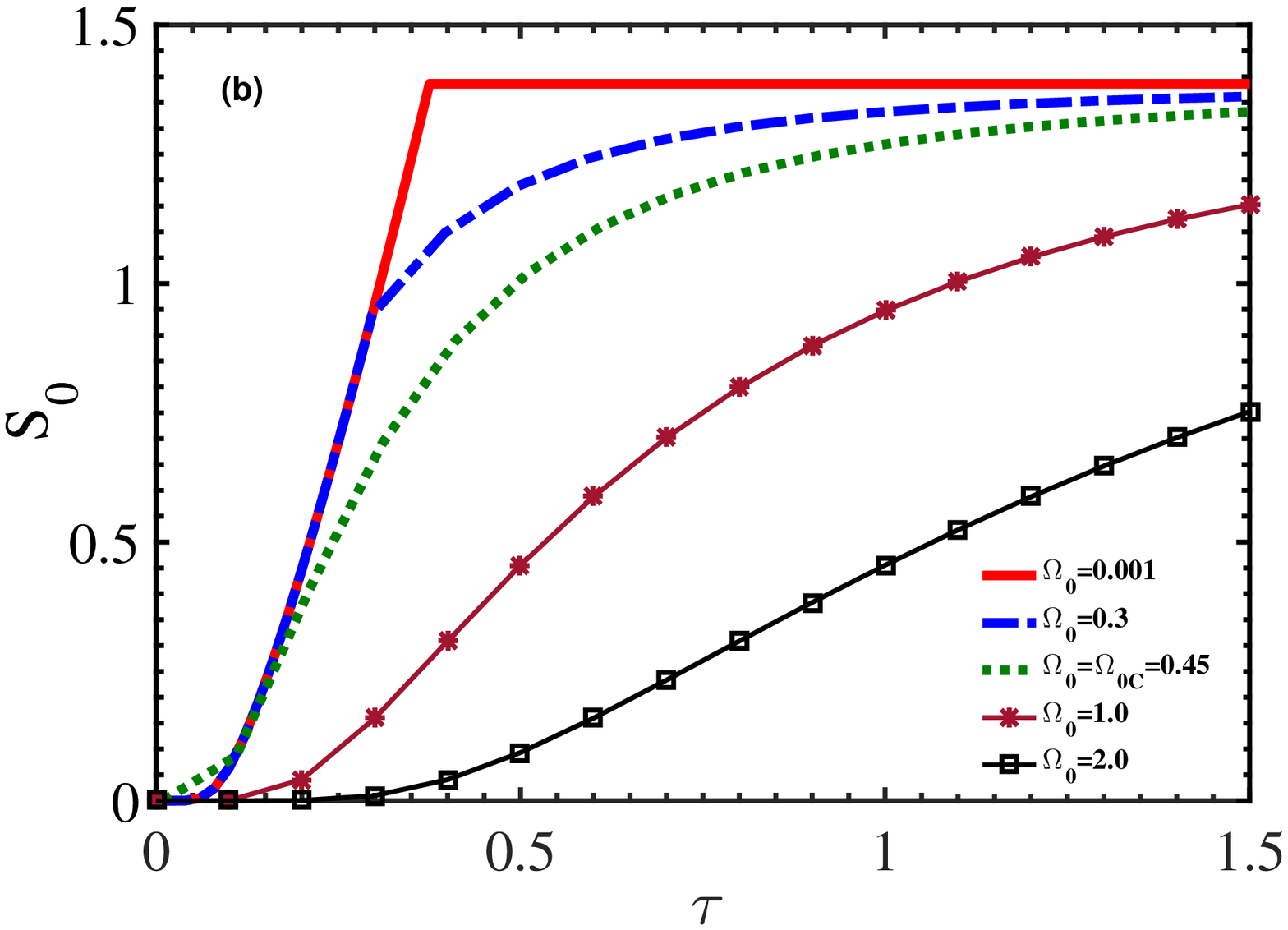} 
	\caption{The temperature dependence of the mean field magnetic entropy S$_0$ for the spin (a) $\rm s=1/2$ and (b) $\rm s=3/2$ cases. We choose $\rm J(0) = 1.2$ for $\rm s=1/2$ and $\rm J(0)=0.3$ for spin $\rm s=3/2$ systems. The values of critical field $\rm\Omega_{0c}$ are given in the figure.}	
\end{figure}

Fig.~1 shows the MFA temperature dependence of the total magnetic moment per site m and its components m$_z$, m$_x$ for the spin-1/2 and the spin-3/2 cases. The spin system is in the QPa state if $\tau> \rm \tau_R$, where the only magnetic component along the transverse field m$\rm_x$ exists. The Curie temperature $\rm \tau_C$ determines an order-disorder phase transition without the external TrF, $\rm \tau_C = \rm \tau_R(\Omega_0=0)$. The critical temperature used in the TIM is practically identical with the spin reorientation temperature $\rm \tau_R$, which reduces with increasing TrF and $\rm \tau_R(\Omega_0) \le \tau_C$. In addition, both the magnetic moment and the spin orientation temperature are enhanced with the increase of the spin value s. 

\begin{figure}[tp]
	\centering
	\includegraphics[height=2.2in,width=2.4in]{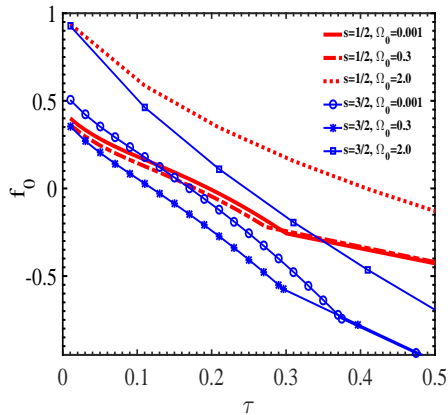}  
	\caption{The free energies f$_0$ of the HS and LS systems versus elevating temperature at different transverse fields $\Omega_0$. The exchange interaction parameters are chosen as in Fig.~3.}	
\end{figure}

The dependence of the spin orientation temperature on the TrF is illustrated in Fig.~2. The $\rm \tau_R$ value and the critical transverse field $\rm \Omega_{0c}$ are proportional to the internal exchange parameter J(0) for both two spin cases. However, for the same set of parameters, the $\rm \tau_C$ value for the HS case is always considerably larger than that for the LS case. At a given Curie temperature $\rm \tau_C$, J(0) will be chosen smaller for the HS model than for the LS model since $\rm \tau_C=\rm J(0)s(s+1)/3$  within the MFA. Consequently, J(0) is set to be 1.2 and 0.3 for the LS and HS cases in Fig.~3-6, respectively. 
\begin{figure*}[tp]
	\centering
	\includegraphics*[height=2.2in,width=2.5in]{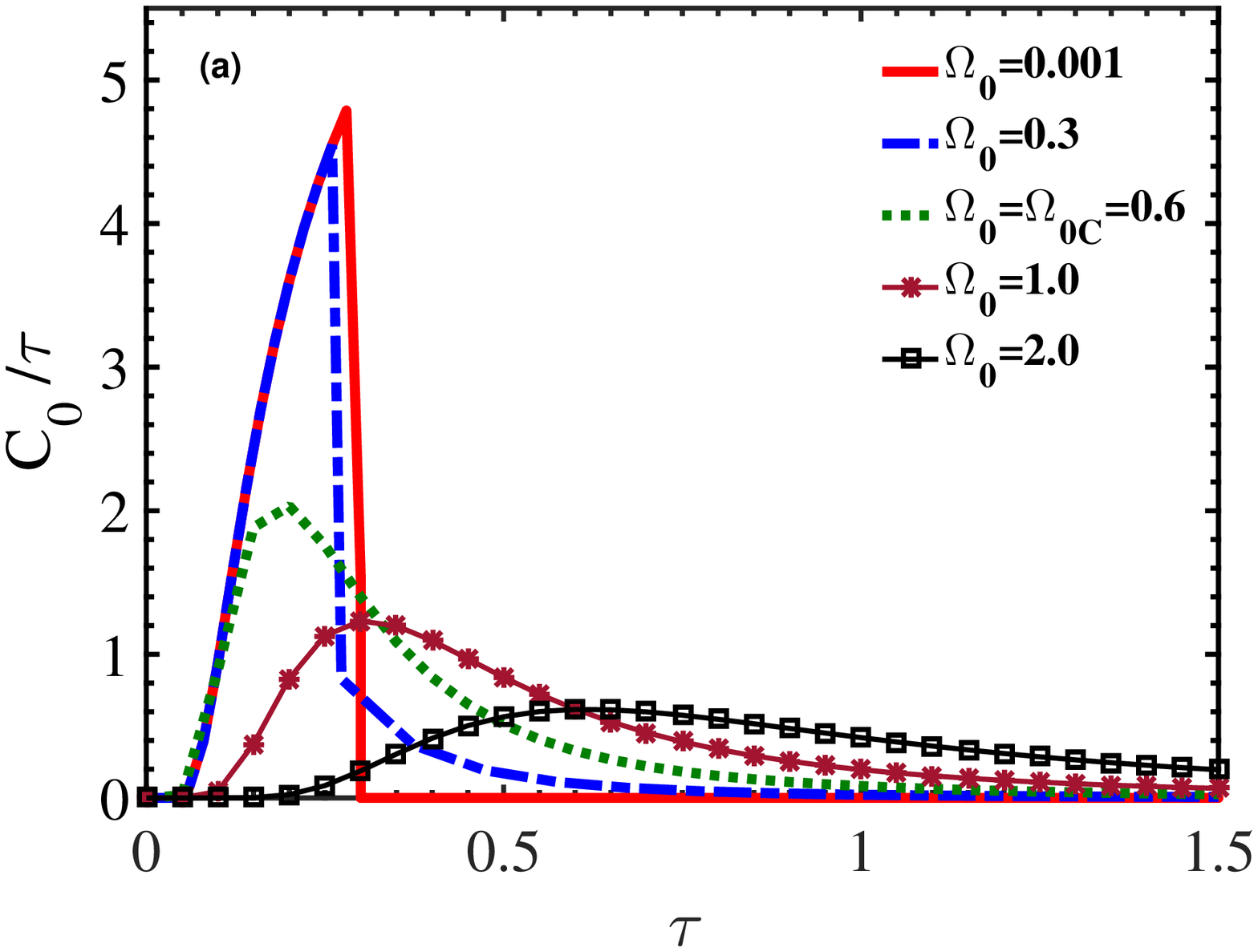}  
	\includegraphics*[height=2.2in,width=2.5in]{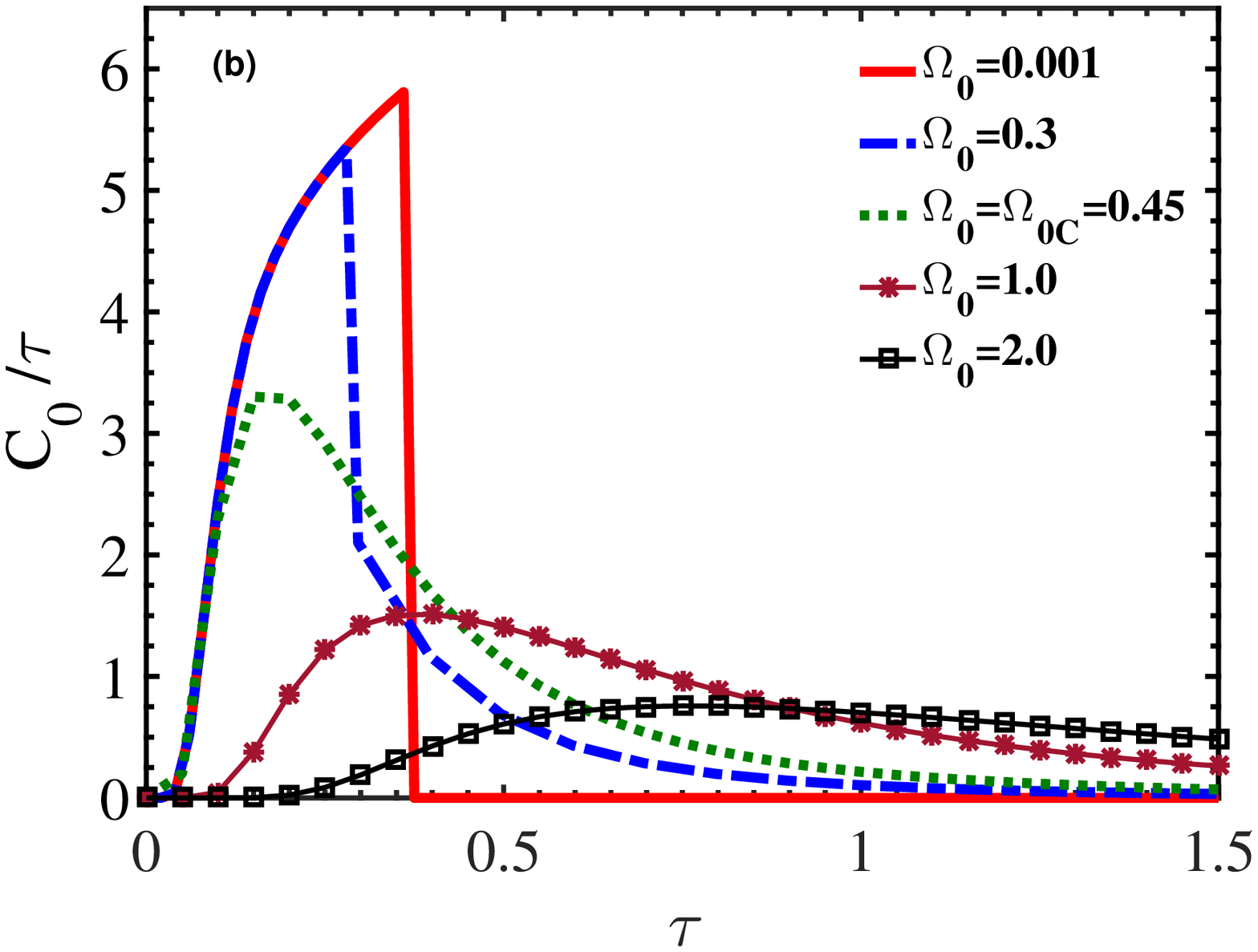}  
	\caption{The dependence of the ratio $\rm C_0/\tau$ on temperature for the  LS (a) s=1/2 and HS (b) s=3/2 systems for different transverse fields. The exchange interaction parameter J(0) is chosen similarly as in Fig.~3.}
\end{figure*}

\begin{figure}[tp]
	\centering
	\includegraphics*[height=2.2in,width=2.5in]{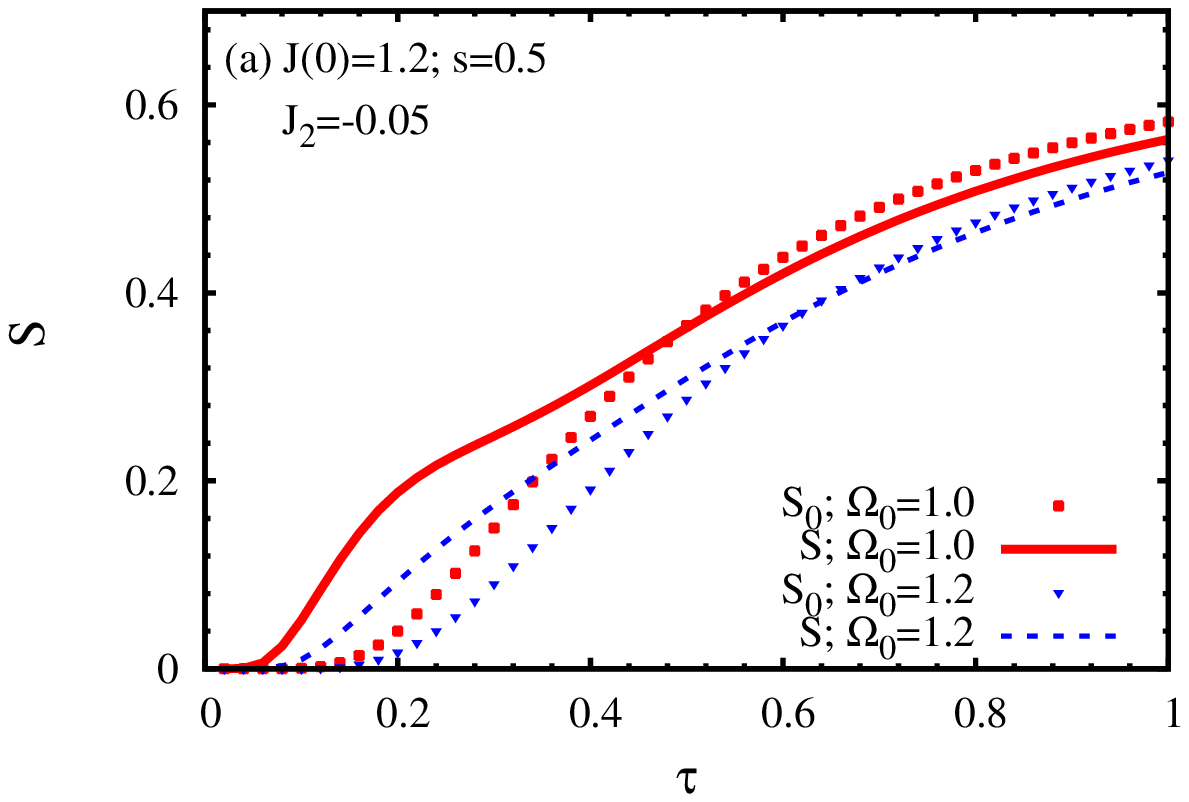}  
	\includegraphics*[height=2.2in,width=2.5in]{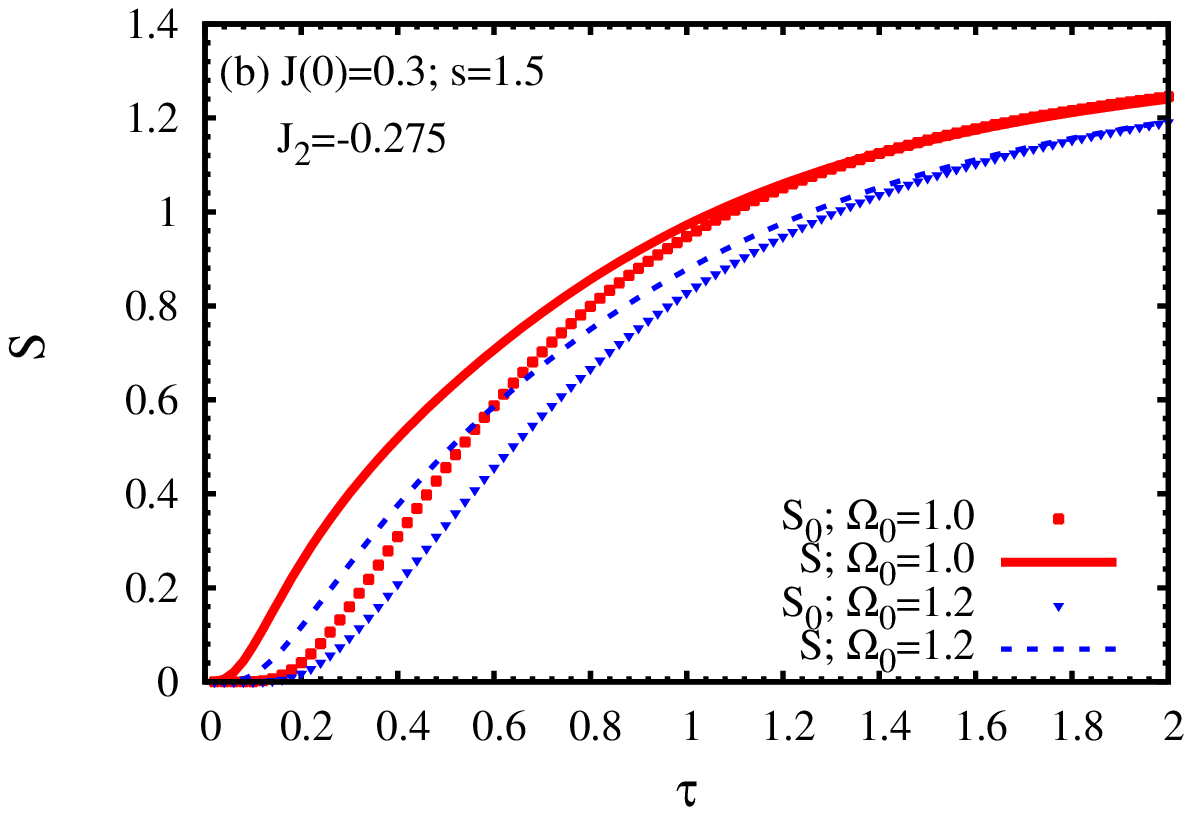}  
	\caption{Temperature dependence of the entropy of the spin systems in the QPa phase with $\Omega_0>\rm \Omega_{0c}$. The MF part and the total entropy including fluctuations are denoted by S$_0$ and S, respectively. The exchange parameters are $\rm J(0)=1.2$, $\rm J'_z=-0.2$, J$_1=-0.1$, J$_2=-0.05$ for the LS (a) and $\rm J(0)=0.3$, $\rm J'_z=-0.2$, J$_1=-0.1$, J$_2=-0.275$ for the HS (b) cases, respectively. The critical TrF value in the MFA is 0.6 (0.45) for the LS (HS) system.}
\end{figure}

We plot in Fig.~3 the spin entropy S$_0$ as a function of temperature at different TrFs for the two spin cases where the entropy is monotonically suppressed with increasing TrF and reaches the saturation values, ln2 for $\rm s=1/2$ and ln4 for $\rm s=3/2$ at high temperatures. The entropy data extracted from experimental results for CoNb$_2$O$_6$ generally agreed with this trend. However, the spin entropy of CoNb$_2$O$_6$ in Fig.~4b of Ref.~\cite{Tian2015} behaves unexpectedly larger at $\rm B=5$ T than at zero field. Typically, the spin entropy of both HS and LS systems at the same temperature must be smaller in larger TrFs (Fig.~3). Therefore, the entropy data at $\rm B=5$ T of Ref.~\cite{Tian2015} are seemingly peculiar, which requires further experimental verification.

Fig.~4 exhibits the temperature-dependent free energy of the LS and HS systems at different TrFs. Apparently, the HS system is more stable than the LS system at high TrFs and at low temperatures due to its lower free energy (see the curves for s $=1/2$ and $\rm s=3/2$ when $\Omega_0=2.0$ with $\tau < 0.5 $). This observation is reasonably expected because the Zeeman energy dominates at high fields and strongly reduces the free energy. This fact is essential to explain the experimental results.

Fig.~5 presents the ratio of the MF heat capacity $\rm C_0/\tau$ and temperature for different spin values. While increasing the TrF $\Omega_0$ from zero to the critical value $\rm \Omega_{0c}$, the maximum peak shifts toward the lower temperature. When $\Omega_0 >\rm \Omega_{0c}$, the spin system exists in the QPa state at zero temperature. In this region, the peak moves to the higher temperature with increasing transverse fields. The shift of the peak of the $\rm C_0/\tau$ curve in the QPa state is experimentally observed in Ref.~\cite{Tian2015} but its nature has not been unveiled. We believe that the maximum peak originates from two opposite tendencies where the transverse field enhances the transverse order and the thermal fluctuations suppress it. Although the MFA result is kindly simple, it presents precisely the qualitative behavior of the $\rm C_0/\tau$ curve.

\subsection{Quantum para-magnetic states within Gaussian approximation}

We are interested in the QPa states when the longitudinal component of the order parameter m$\rm_z$ disappears and the system is completely characterized by the transverse order parameter m$_x=b_s(\Omega_0/\tau)$ (see Eq.~(\ref{Eq23})).  Within the Gaussian approximation, free energy f is 								
\be
\rm f = \rm f_0 + {1 \over 2\beta N} \sum_{\bf{k}} \rm ln {sh( \beta\omega_{\bf k}/2) \over sh(y_0/2)}\,,
\label{Eq24}
\ee
where
\be
\rm f_0 = -{1\over \beta} \rm ln {sh[(s+1/2)y_0] \over sh(y_0/2)}.
\ee 
 
The entropy is given by
\bea
\rm S &=& \rm S_0 -{1\over 2N} \sum_{\bf{k}} ln {sh(\beta\omega_{\bf k}/2) \over sh(y_0/2)} \nonumber \\ &+& {1\over 4N} \sum_{\bf{k}} \Big\{ [\beta\omega_{\bf{k}}-\beta \rm J({\bf{k}})y_0b'_s(y_0)] cth(\beta\omega_{\bf{k}}/2)-y_0cth(y_0/2) \Big\} \,,
\eea
where
\be
\rm S_0 = \rm -y_0b_s(y_0)+ln {sh[(s+1/2)y_0] \over sh(y_0/2)}.
\ee
 
The specific heat is 
\be
\rm C = \rm C_0 + {y^2_0 \over 8sh^2(y_0/2)} + {\beta^2 \over 8N} \sum_{\bf{k}} {[\omega_{\bf{k}}-J({\bf{k}})y_0b'_s(y_0)]^2 \over sh^2({\beta\omega_{\bf{k}}/2})},\,\,\,\,\,\,\,\,\,\,\,
\label{Eq27}
\ee
where $\rm C_0= \rm y^2_0 b'(y_0)$ and the elementary excitation energy in the QPa state is 
\bea
\omega_{\bf{k}}&=&\rm \Omega_0-J({\bf{k}})b_s(y_0),\\
\rm y_0&=&\beta\Omega_0.
\label{Eq28}
\eea

Elementary excitations at finite temperature contribute to the additional second term {in Eq.~(\ref{Eq24}) for free energy beyond the MFA. Therefore, the thermodynamic properties of the spin system are calculated numerically using the Eqs.~(\ref{Eq24})-(\ref{Eq28}) where the summation taken over $\bf k$ values is replaced by the integration,
\be
\rm {1 \over N} \sum_{\bf k} ... \rightarrow {1\over ({4\pi})^3} \int_{-2\pi}^{2\pi} dk_x \int_{-2\pi}^{2\pi} dk_y \int_{-2\pi}^{2\pi} dk_z \nonumber\,\,\,,
\ee
and
\be
\rm J({\bf k})=2[\rm{J_zcos(k_z/2)+J'_zcos(k_z)+J_1cos(k_y)}+\rm{2J_2cos(k_x/2)cos(k_y/2)]}
\label{Eq29} .
\ee
Fig.~6 shows the temperature dependence of the MFA entropy S$_0$ and the total fluctuating entropy S of the LS and the HS systems in the QPa states at different TrFs. At very low temperatures, because of the Heisenberg uncertainty principle, the quantum spin fluctuations have stronger influence on the entropy than the thermal fluctuations. An increase in TrF $\Omega_0$ enhances the order parameter m$\rm _x$ in the QPa states, thus reduces the disorder and the spin entropy. At the same field near the critical field, a bump of the entropy curve is more visible at very low temperatures for the LS case and is suppressed with increasing TrFs (see Fig.~6a). 

\begin{figure}[tp]
	\centering
	\includegraphics[height=2.4in,width=2.6in]{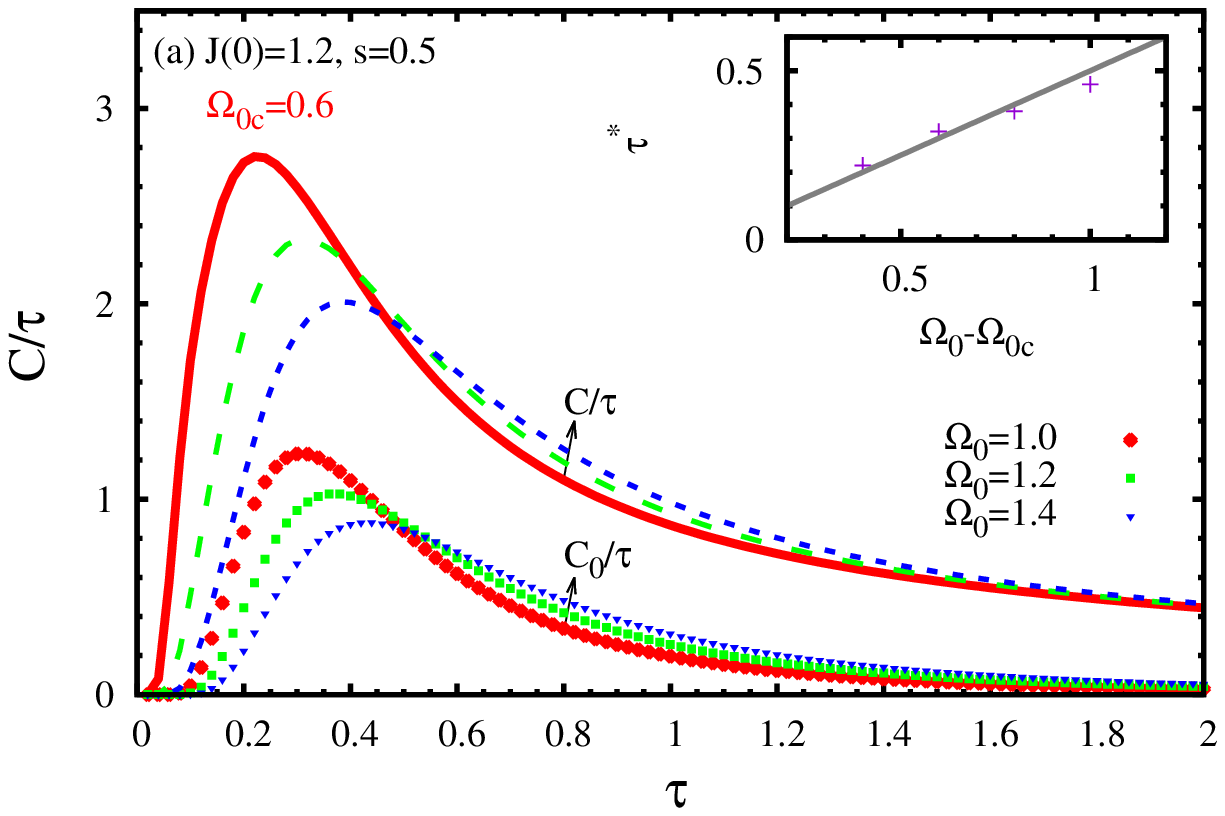}  
	\includegraphics[height=2.4in,width=2.6in]{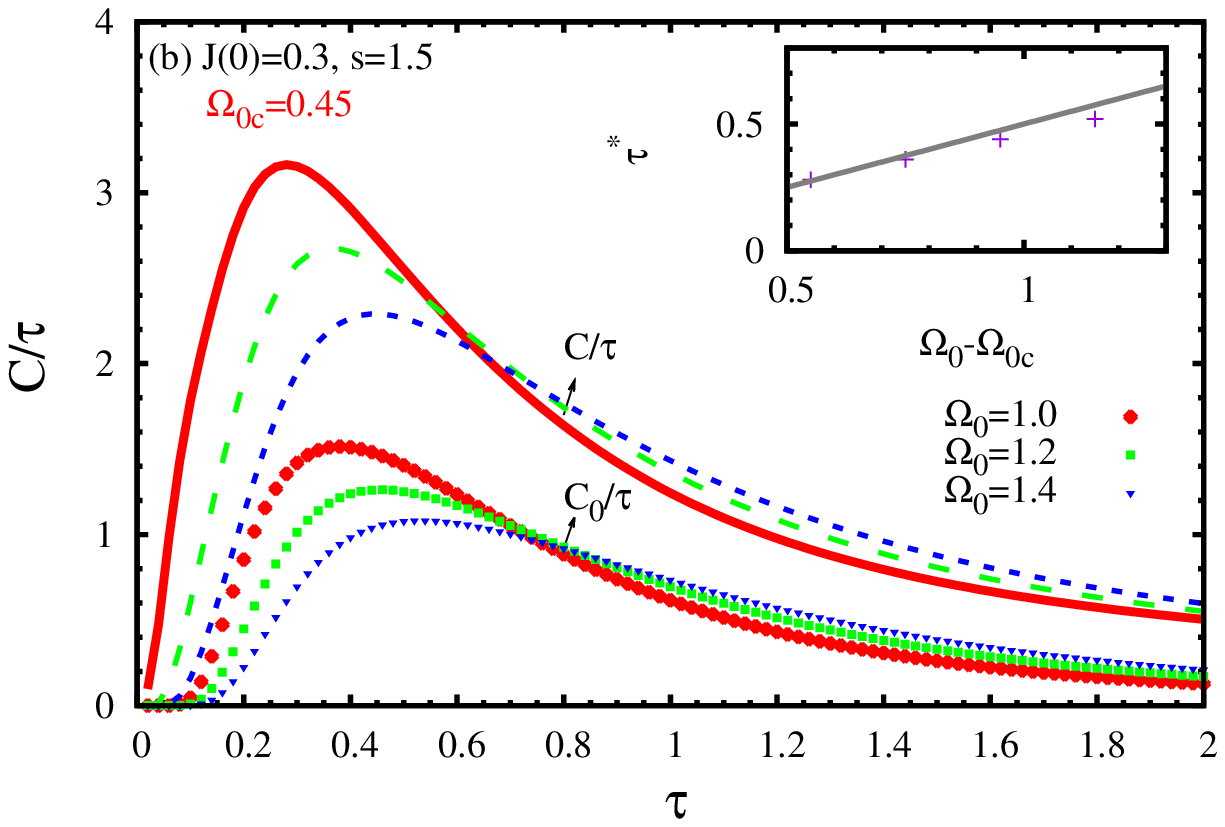}  
	\caption{The temperature dependence of $\rm C/\tau$ in the QPa phase with $\Omega_0>\Omega_{0c}$. The MF part and the total entropy including fluctuations, are denoted by C$_0$ and C. The exchange parameters J(0), $\rm J'_z$, J$_1$, J$_2$ for the LS (a) and for the HS (b) cases are chosen similarly to Fig.~6. The insets show the linear dependence of the characteristic temperature $\tau*$ on the field difference $\Omega_0-\Omega_{0c}$.}
\end{figure}

In order to emphasize the influence of the spin fluctuations on $\rm C/\tau$, we show both $\rm C/\tau$ and $\rm C_0/\tau$ with and without the contribution of the spin fluctuations in Fig.~7. We can assess the spin fluctuation effect from the deviation of the specific heat from its MFA value, i.e. $\Delta \rm C=C-C_0$. The spin fluctuations strongly affect the spin systems in the low temperature regime, which is characterized by the enhancement of the $\rm C/\tau$ peak near zero temperature. The amplitude of this peak is larger when the TrF is closer to the critical field $\rm \Omega_{0c}$. The temperature $\tau^*$ corresponding to the maximum can be estimated at the zero temperature limit of Eq.~(\ref{Eq27}). The analytic calculations show that the specific heat tends to zero following by the exponential law exp$[-(\Omega_0-\rm \Omega_{0c})/\tau]/\tau^2$ and the maximum of $\rm C/\tau$ curve occurs at $\tau^* \approx \rm (\Omega_0-\Omega_{0c})/2$ with $\rm \Omega_{0c}=J(0)s$. The insets in Fig.~7 clearly describe the linear dependence of $\tau^*$ on the deviation from the critical transverse field $\rm \Omega_{0c}$ . 

\begin{figure}[tp]
	\centering
	\includegraphics[height=2.4in,width=2.6in]{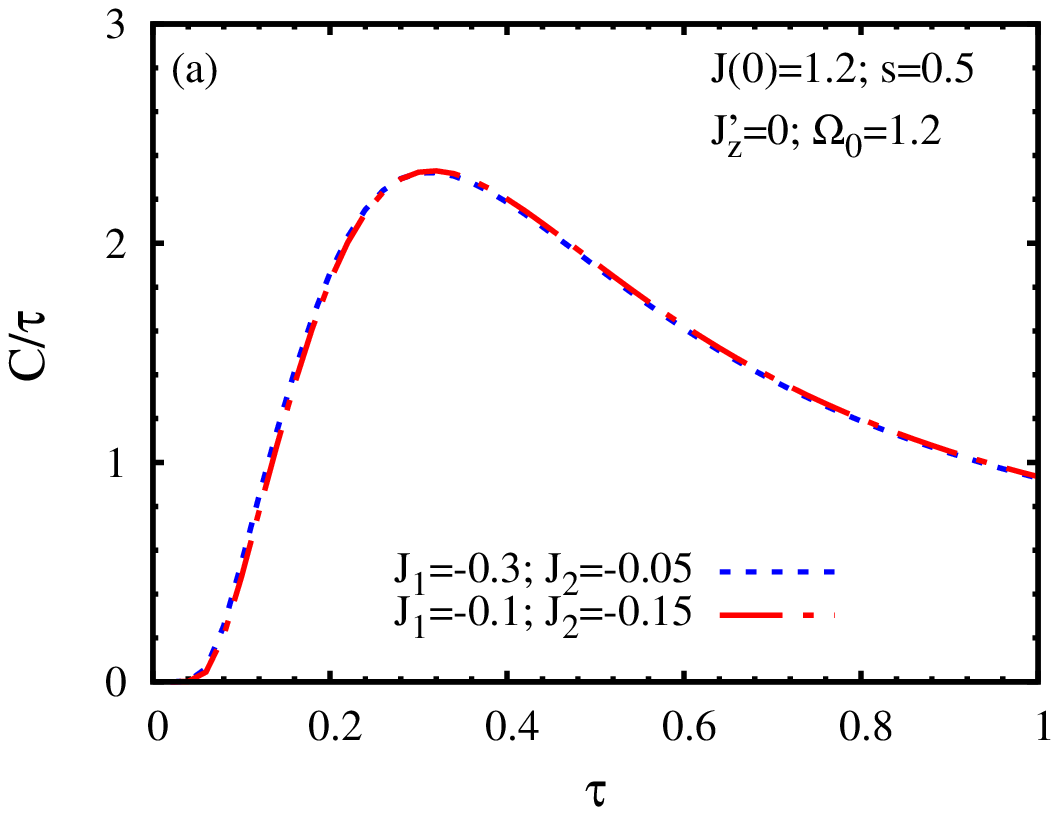}  
	\includegraphics[height=2.4in,width=2.6in]{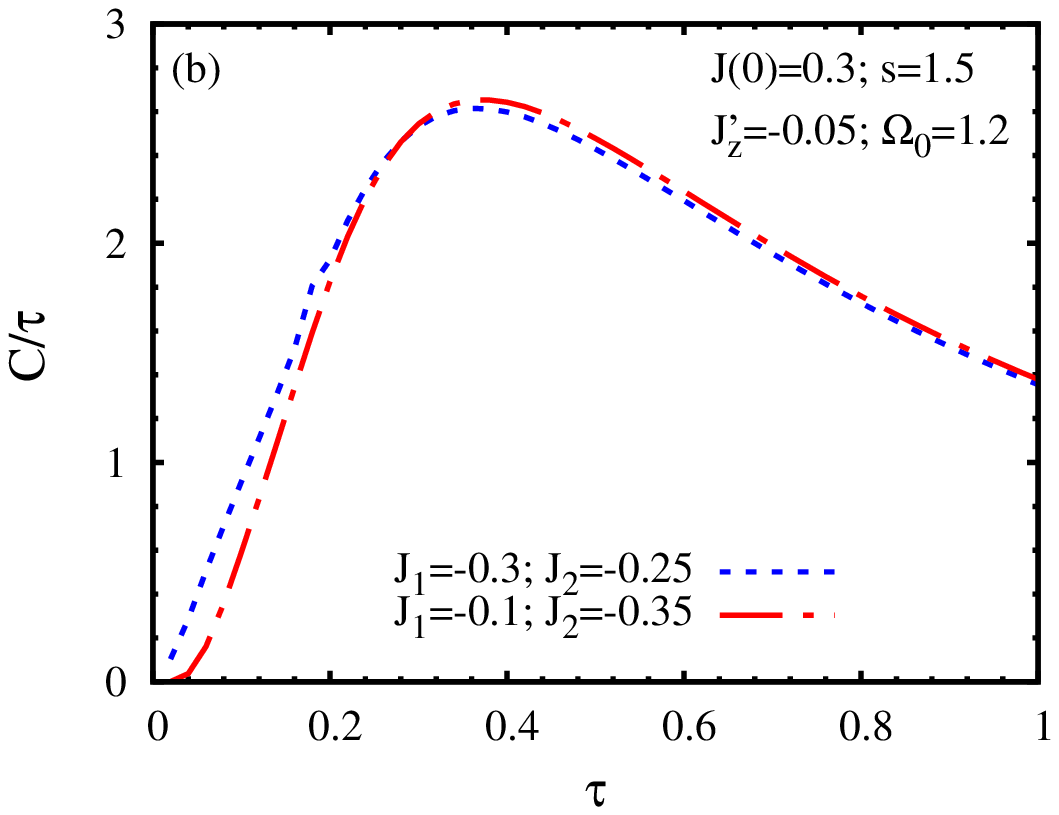}  
	\caption{The dependence of $\rm C/\tau$ on temperature $\tau$ with various NN in-plane anti-ferromagnetic exchange and  NNN  intra-chain exchange integrals for spins (a) $\rm s=1/2$ and (b) $\rm s=3/2$. The TrF is $\Omega_0=1.0$.}
\end{figure}

We next investigate the modification of specific heat on the in-plane inter-chain couplings. Fig.~8(a) shows that in the LS case, the anti-ferromagnetic NN inter-chain exchange couplings, J$_1$, J$_2$, slightly affect the shape and the magnitude of the temperature dependence of the heat capacity at the same sufficiently large value J(0)=1.2. In the HS case, a significant change and a shift of the maximum peak of the $\rm C/\tau$ curve to the lower temperature are observed when J$_1$ and J$_2$ values are comparable with the exchange parameter J(0)=0.3 (see Fig.~(8b)). In the HS 3D-TIM, the AF inter-chain in-plane exchange couplings J$_1$ and J$_2$ play a key role in the formation of the isosceles triangular spin lattice and they noticeably affect the heat capacity near the critical temperature.

\section{The specific heat of CoNb$_2$O$_6$ in the quantum para-magnetic states}

In this part, the thermodynamic properties of a typical 1D Ising ferro-magnet CoNb$_2$O$_6$ in the QPa states are numerically calculated and are discussed in the framework of the Gaussian spin fluctuation approximation.

The phase transition temperature and the critical TrF derived from the field dependent specific heat experiment \cite{Tian2015} are about $\rm T_c=2.85$ K or 0.246 meV and $\rm B_c=5.24$ T or 0.61 meV, respectively. One can use these data to estimate the order of the exchange coupling parameter J(0) defined by Eq.~(\ref{Eq29}). Within the MFA, the Curie temperature $\rm T_c$ is evaluated by $\rm J(0)s(s+1)/3$. Taking $\rm T_c=2.85 K$, we obtain the exchange parameter $\rm J(0)=1.039$ meV for the LS case and $\rm J(0)=0.197$ meV for the HS case.  Using the best fit for the experimental curves at $\rm B = \rm 5.4 \,\,T, 6.5 \,\,T, 8 \,\,T $ (see \cite{Tian2015}) and taking into account that the Curie temperature is normally overestimated by the MFA, we derive the exchange parameter values listed in the last two rows of Table 1.

%\begin{figure}[tp]
%	\centering
%	\includegraphics[height=2.8in,width=3.in]{Fig_9.eps}  
%	\caption{ Experimental and theoretical temperature dependence of the specific heat data of CoNb$_2$O$_6$ at 5.4 T. The dashed (solid) line corresponds to the 1D (3D) TIM with exchange coupling parameters given in Table 1.}
%\end{figure}

\begin{figure}[tp]
	\centering
	\includegraphics[height=4.2in,width=3.2in]{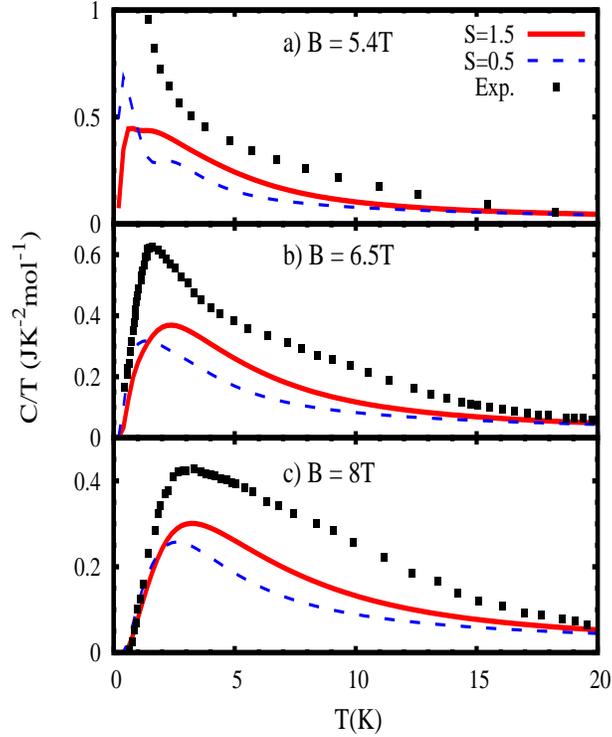}  
	\caption{The temperature dependence of the experimental specific heat data \cite{Tian2015} at $\rm B = 5.4$ (a), 6.5 (b) and 8 T (c) by the LS (dashed lines) and the HS (solid lines) 3D-TIM. The exchange coupling parameters of the models are given in the last two rows of Table 1. }
\end{figure}

\begin{table}[]
	\caption{Exchange coupling parameters (meV)}
	\begin{tabular}{|c|cccccc|}
		\cline{1-7}
		{\bf TIM}	& \multicolumn{6}{c|}{{\bf Exchange coupling parameters (meV)}} \\ \cline{1-7} 
		\multirow{5}{*}{\bf S=0.5} & $\rm J_z$  & $\rm J'_z$   & $\rm J_1$  & $\rm J_2$  & $\rm J(0)$   & Ref.  \\ \cline{2-7}
		& 2.19  & -0.29  & -0.03  & -0.02  &  3.66 & [12]  \\
		& -0.152  & 0.332  & -0.106  & -0.280  & $\rm NA$  & [9]  \\
%		& 0.519  & 0  & 0  & 0  & 1.039  & Current work 1D-TIM  \\
		& 0.958  & -0.043  & -0.043  & -0.172  & 1.056  & Current work 3D-TIM \\ \cline{1-7}
		\bf S=1.5 & 0.362  & -0.043  & -0.086  & -0.043   & 0.294  & Current work 3D-TIM \\
		\cline{1-7}
	\end{tabular}
\end{table}

Table 1 also lists the exchange coupling parameters $\rm J_z$, $\rm J'_z$, $\rm J_1$, $\rm J(0)$ of the TIM for the CoNb$_2$O$_6$ spin system in the QPa states which are extracted from neutron experiments \cite{Cabrera2014} at 7 T and from the density functional theory (DFT) calculation \cite{Molla2018}. Comparing data given in Table 1, we note that the value $\rm J(0)=3.66$ meV estimated from the neutron experiment \cite{Cabrera2014} is about four times for the LS model and is twelve times for the HS model larger than the values evaluated within our theory. The signs of intra-chain couplings $\rm J_z$ and $\rm J'_z$ obtained by the DFT calculations \cite{Molla2018} seem to be opposite to what have been used in Ref.~\cite{Cabrera2014}. However, the magnitudes of these exchange parameters reasonably agree with the magnitudes of the parameters extracted from the specific heat measurements. The temperature dependence of the specific heat of CoNb$_2$O$_6$ has been investigated when $\rm B>B_c$ (5.24 T) in Ref.~\cite{Tian2015}, which shows that the spin system exists in the QPa at zero temperature.

Fig.~9 exhibits the fit using the 3D-LS and -HS TIM for the specific heat data of CoNb$_2$O$_6$ at $\rm B\, =\,5.4$, 6.5 and 8 T. The specific heat behavior near zero temperature is better described by the 3D-LS model but the 3D-HS model is quantitatively closer to the experimental values at the high temperature. Since the low spin model (s=1/2) is more "quantum", it is more appropriate to describe the thermodynamic properties near zero temperature. At elevated temperatures and at high fields, the spin-3/2 model is more adequate to explain the behaviors of CoNb$_2$O$_6$ since the high energy excited states make a significant contribution to the thermodynamic properties. We note that a spin crossover from low to high spin state is possible with increasing transverse field. The influence of the spin crossover on the thermodynamics of the ferroics is an intriguing subject for further study.  

\section{Conclusions}

The thermodynamics of ferroics having quantum phase transition are examined using the TIM with different spins in the framework of the mean field and the Gaussian spin fluctuation approximations. The 3D-TIM model with various spin values successfully illustrates the suppression and the shift of the specific heat maximum in the QPa states experimentally observed in CoNb$_2$O$_6$. The peak of the specific heat in the QPa phase near zero temperature is vividly described by using the 3D-TIM with spin-1/2. However, the spin-3/2 3D-TIM is more suitable to present the temperature dependent specific heat of CoNb$_2$O$_6$ in the QPa state at high fields with increasing temperature. 

\section*{Acknowledgment}

The authors thank NAFOSTED Grant No.~103.01-2015.92 for support.



%%       \citet{<label>} ==> Jones et al. [21]
%%       \citep{<label>} ==> [21]
%%


\section*{References}
\begin{thebibliography}{9}
	
	\bibitem{Scott2015} J.~F.~Scott, A.~Schilling, S.~E.~Rowley, and J.~M.~Gregg, Sci.~Tech.~ Adv.~Mater.~{\bf 16}, 036001 (2015).
	
	\bibitem{Rowley2014} S.~E.~Rowley, L.~J.~Spalek, R.~P.~Smith, M.~P.~M.~Dean, M.~Itoh, J.~F.~Scott, G.~G.~Lonzarich and S.~S.~Saxena, 
	Nat.~Phys.~{\bf 10}, 367 (2014).
	
	\bibitem{Sachdev1999} S.~Sachdev, {\it Quantum Phase Transitions} (Cambrige University Press, Cambrige, England, 1999).
	
	\bibitem{Kinross2014} A.~W.~Kinross, M.~Fu, T.~J.~Munsie, H.~A.~Dabkowska, G.~M.~Luke, Subir Sachdev, and T.~Imain, Phys.~Rev.~X {\bf 4}, 031008 
	(2014).
	
	\bibitem{Tian2015} Tian Liang, S.~M.~Koohpayeh, J.~W.~Krizan, T.~M.~McQueen, R.~J.~Cava, N.~P.~Ong, Nat.~Commun.~{\bf 6}, 7611 (2015). 
	
	\bibitem{Lieb1961} E.~H.~Lieb, T.~D.~Schultz, D.~C.~Mattis, Ann.~Phys.~{\bf 16}, 407 (1961).
	
	\bibitem{Pfeuty1970} P.~Pfeuty, Ann.~Phys.~{\bf 57}, 79 (1970).
	
	\bibitem{Hanawa1994} T.~Hanawa, K.~Shikawa, M.~Ishikawa, K.~Miyatani, K.~Saito and K.~Kohn, J.~Phys.~Socie.~Jpn. {\bf 63}, 2706 (1994).
	
	\bibitem{Molla2018} K.~Molla, B.~Rahaman, in {\it AIP Conference Proceeding 1953}, 120011 (2018).
	
	\bibitem{Cong2018} Niem T.~Nguyen, Thao H.~Pham, Giang H.~Bach, Cong T.~Bach, Mater.~Trans.~{\bf 59}, 1075 (2018).
	
	\bibitem{Suzuki2013} S.~Suzuki, J.~Inoue, B.~K.~Chakrabarti, {\it Quantum Ising Phases and Transitions in Transverse Ising Models} (Springer, 
	2013, 2nd ed.). 
	
	\bibitem{Cabrera2014} I.~Cabrera, J.~D.~Thompson, R.~Coldea, and D.~Prabhakaran, Phys.~Rev.~B {\bf 90}, 014418 (2014).
	
	\bibitem{Kane2004} T.~Kaneyoshi, Physica A {\bf 339}, 403-415 (2004).
	
	\bibitem{Heid1995} C.~Heid, H.~Weitzel, P.~Burlet, M.~Bonnet, W.~Gonschorek, T.~Vogt, J.~Norwig, and H.~Fuess, J.~Magn.~Magn.~Mater.~{\bf 151}, 123 (1995). 
	
	\bibitem{Cong2016} Nguyen Tu Niem, Bach Huong Giang, Bach Thanh Cong, J.~Sci.~Adv.~Mater.~Devices {\bf 1}, 531 (2016).
	
	
\end{thebibliography}


\end{document}